\documentclass[preprint,nofootinbib,aps,superscriptaddress]{revtex4}
\usepackage{color,graphicx,shortvrb,epsfig}

\usepackage{color,shortvrb,epsfig}

\newcommand{\pslinear}[5][]{\psline [#1]#2#3 \psline [#1]{#4}#2#5}
\newcommand{\wt}{\widetilde}

\usepackage{graphicx,hyperref}
\newcommand{\beq}{\begin{equation}}
\newcommand{\eeq}{\end{equation}}
\newcommand{\beqa}{\begin{eqnarray}}
\newcommand{\eeqa}{\end{eqnarray}}
\newcommand{\no}{\nonumber}
\def\OMIT#1{{}}
\newcommand{\lsim}{\mathrel{\rlap{\lower4pt\hbox{\hskip1pt$\sim$}}
    \raise1pt\hbox{$<$}}}         %less than or approx. symbol
\newcommand{\gsim}{\mathrel{\rlap{\lower4pt\hbox{\hskip1pt$\sim$}}
    \raise1pt\hbox{$>$}}}         %greater than or approx. symbol
\newcommand{\epss}{\epsilon_{\rm S}}

\begin{document}

\preprint{\vbox{\hbox{}\hbox{}\hbox{} \hbox{SLAC-PUB-10519}
    \hbox{WIS/18/04-July-DPP} \hbox{hep-ph/0407063}}}

\vspace*{1.5cm}

\title{\boldmath New Ways to Soft Leptogenesis}

\author{Yuval Grossman}\email{yuvalg@physics.technion.ac.il}
\affiliation{Department of Physics, Technion--Israel Institute of
  Technology, Technion City, 32000 Haifa, Israel}
\affiliation{Stanford Linear Accelerator Center, 
Stanford University, Stanford, CA 94309}
\affiliation{Santa Cruz Institute for Particle Physics,
University of California, Santa Cruz, CA 95064}

\author{Tamar Kashti}\email{tamar.kashti@weizmann.ac.il}
\affiliation{Department of Particle Physics,
  Weizmann Institute of Science, Rehovot 76100, Israel}

\author{Yosef Nir}\email{yosef.nir@weizmann.ac.il}
\affiliation{Department of Particle Physics,
  Weizmann Institute of Science, Rehovot 76100, Israel}

\author{Esteban Roulet}\email{roulet@cab.cnea.gov.ar}
\affiliation{CONICET, Centro At\'omico Bariloche,
  Av. Bustillo 9500, 8400 Bariloche, Argentina}

%\date{\today}
%\pacs{12.10.Dm, 12.10.Kt, 98.80.Cq}

\vspace{2cm}
\begin{abstract}
Soft supersymmetry breaking terms involving heavy singlet sneutrinos
provide new sources of lepton number violation and of CP
violation. In addition to the CP violation in mixing, investigated
previously, we find that `soft leptogenesis' can be generated by CP
violation in decay and in the interference of mixing and decay. These
additional ways to leptogenesis can be significant for a singlet
neutrino Majorana mass that is not much larger than the supersymmetry
breaking scale, $M\lsim10^2 m_{\rm SUSY}$. In contrast to CP violation
in mixing, for some of these new contributions the sneutrino
oscillation rate can be much faster than  the decay rate, so that the
bilinear scalar term need not be smaller than its natural scale.  
\end{abstract}

\maketitle

%%%%%%%%%%%%%%%%%%%%%%%%%%%%%%%%%%%%%%%%%%%%%%%%%%%%%%%%%%%%%%%%%%%%%%
\section{Introduction}
\label{sec:introduction}
The evidence for neutrino masses at a scale $\sim10^{-2}$ eV makes a
convincing case for the seesaw mechanism
\cite{Gell-Mann:vs,yanagida,Mohapatra:1979ia}: The existence of singlet
neutrinos with Majorana masses and with Yukawa couplings to active
neutrinos becomes very plausible. The physics of these heavy neutrinos
can provide all the necessary ingredients for baryogenesis
\cite{Sakharov:dj}: $B-L$ is violated by the Majorana masses, CP 
is likely to be violated in the neutrino Yukawa couplings and, for
small enough Yukawa couplings, the heavy neutrinos decay out of
equilibrium. Thus, leptogenesis \cite{Fukugita:1986hr}, the dynamical
generation of lepton asymmetry through the decays of heavy singlet
Majorana neutrinos, becomes an attractive solution to the
puzzle of the baryon asymmetry.

The seesaw mechanism introduces a new scale, $M$, the mass scale
of the singlet neutrinos. Since this scale must be much higher
than the electroweak breaking scale, $M\gg\Lambda_{\rm EW}$, a
huge amount of fine-tuning is required within the framework of the
Standard Model extended to include singlet neutrinos (SM+N) to
keep the low Higgs mass. This situation provides further
motivation to consider the supersymmetric extension of the model
(SSM+N). Then, leptogenesis is induced in both singlet neutrino
and singlet sneutrino decays. The results are modified by factors
of order one, but the basic mechanism and the order of magnitude
of the asymmetry remain very much the same as in the
non-supersymmetric version.

Supersymmetry must, however, be broken. In addition to the soft
supersymmetry breaking terms of the SSM, there are now terms that
involve the singlet sneutrinos $\widetilde N$, in particular, 
bilinear ($B$) and trilinear ($A$) scalar couplings. These terms
provide yet another source of lepton number violation and of CP
violation. One may ask whether these terms can play a significant role
in leptogenesis. One finds that for a certain range of parameters, the
soft breaking terms play a significant role, and may even be dominant
in leptogenesis \cite{Grossman:2003jv,D'Ambrosio:2003wy}. This
scenario has been termed `soft leptogenesis.' (For related work, see 
\cite{Giudice:2003jh,Hambye:2004jf,Chun:2004eq,Allahverdi:2004ix,Boubekeur:2004ez}.)

In \cite{Grossman:2003jv} we investigated soft leptogenesis
related to CP violation in mixing (a leptonic analog of
R$e(\epsilon)_{K\to\pi\ell\nu}$).  In this work, we present all the
contributions to the lepton asymmetry that arise in this scenario. The
contribution considered in \cite{Grossman:2003jv} dominates over
the other contributions in a large part of the parameter space. Yet, if the
scale $M$ is relatively low, other contributions, related to CP
violation in the interference of decays with and without mixing (a
leptonic analog of $S_{B\to\psi K}$), and to CP violation in decay (a
leptonic analog of R$e(\epsilon^\prime)_{K\to\pi\pi}$), play a
significant role. 

The plan of this paper is as follows. In section \ref{app:mide} we
derive exact expressions for the singlet sneutrino decay rates into
final (s)leptons in terms of mixing and decay amplitudes. In section
\ref{sec:ssm} we present our model, that is the supersymmetric
standard model extended to include singlet neutrinos (SSM+N) and
express the mixing and decay amplitudes in terms of the model
parameters. Our main results are obtained in sections \ref{sec:cpa}
and \ref{sec:siz}. In section \ref{sec:cpa} we evaluate the lepton
asymmetry in terms of the model parameters and, in particular,
assuming hierarchy between the supersymmetry breaking scale and the
mass scale of the singlet sneutrinos, find the potentially leading
contributions. In section \ref{sec:siz} we estimate the size of the
various contributions and find the regions in the SSM+N parameter
space where these contributions can account for the observed baryon
asymmetry. We summarize our results and draw further conclusions in
section \ref{sec:con}. Additional points are made in two
appendices. In appendix \ref{app:thr} we explicitly prove that the
consideration of three body final states does not change the
picture. In appendix \ref{app:smb} we discuss the possibility of
naturally obtaining a small $B$ term for the singlet sneutrinos.

%%%%%%%%%
%%%%%%%%%
\section{Mixing and decay}
\label{app:mide}
We would like to calculate the CP-violating lepton asymmetry:
\beq\label{defeel}
\varepsilon_\ell\equiv\frac{\Gamma(\widetilde L)+\Gamma(L)-\Gamma(\widetilde
  L^\dagger)-\Gamma(\overline L)}{\Gamma(\widetilde L)+\Gamma(L)+\Gamma(\widetilde
  L^\dagger)+\Gamma(\overline L)},
\eeq
where $\Gamma(X)$ is the time-integrated decay rate into a final state
with a leptonic content $X$. Here $L(\overline L)$ is the (anti)lepton
doublet and $\widetilde L(\widetilde L^\dagger)$ is the (anti)slepton
doublet. 

A crucial role in our results is played by the
$\widetilde N-\widetilde N^\dagger$ mixing amplitude,
\beq\label{defmonetwo}
\langle \widetilde N|{\cal H}|\widetilde
N^\dagger\rangle=M_{12}-\frac i2\Gamma_{12},
\eeq
which induces mass and width differences,
\beq\label{defxy}
x\equiv\frac{\Delta M}{\Gamma}\equiv\frac{M_H-M_L}{\Gamma},\ \ \ \ \
y\equiv\frac{\Delta\Gamma}{2\Gamma}\equiv\frac{\Gamma_H-\Gamma_L}{2\Gamma},
\eeq
($\Gamma$ is the average width) between the two mass eigenstates, the
heavy $|\tilde N_H\rangle$ and the light $|\tilde N_L\rangle$,
\beq\label{defpq}
|\widetilde N_{L,H}\rangle=p|\widetilde N\rangle\pm q|\widetilde
N^\dagger\rangle.
\eeq
The ratio $q/p$ depends on the mixing amplitude ratio:
\beq
\left(\frac qp\right)^2=
\frac{2M_{12}^*-i\Gamma_{12}^*}{2M_{12}-i\Gamma_{12}}.
\eeq
For each final state $X$, we define a pair of amplitudes and a
quantity $\lambda_X$ involving the amplitude ratio and the
mixing amplitudes:
\beq\label{decamp}
A_X=\langle X|{\cal H}|\widetilde N\rangle,\ \ \
{\overline A}_X=\langle X|{\cal H}|\widetilde N^\dagger\rangle,\ \ \ 
\lambda_{X}=\frac qp \frac{{\overline A}_{X}}{A_{X}}.
\eeq

Defining $|\widetilde N(t)\rangle$ and
$|\widetilde N^\dagger(t)\rangle$ to be the states that evolve
from purely $|\widetilde N\rangle$ and $|\widetilde
N^\dagger\rangle$, respectively, at time $t=0$, we obtain the
following time-dependent decay rates into a final state $X$:
\beqa\label{tdratestwo}
\Gamma(\widetilde N(t)\to X)&=&{\cal N}_X|A_X|^2e^{-\Gamma t}\left[
  \frac{1+|\lambda_X|^2}{2}\cosh\frac{\Delta\Gamma t}{2}
  +\frac{1-|\lambda_X|^2}{2}\cos(\Delta M\ t)\right.\no\\
&&+\left.{\cal R}e\lambda_X\sinh\frac{\Delta\Gamma
t}{2}-{\cal I}m\lambda_X\sin(\Delta M\ t)\right],\no\\
\Gamma(\widetilde N^\dagger(t)\to X)&=&{\cal N}_X|A_X|^2\left|\frac pq\right|^2
e^{-\Gamma t}\left[
  \frac{1+|\lambda_X|^2}{2}\cosh\frac{\Delta\Gamma t}{2}
  -\frac{1-|\lambda_X|^2}{2}\cos(\Delta M\ t)\right.\no\\
&&+\left.{\cal R}e\lambda_X\sinh\frac{\Delta\Gamma
t}{2}+{\cal I}m\lambda_X\sin(\Delta M\ t)\right],
\eeqa
where ${\cal N}_X$ is a phase space factor.
Summing over the initial states, $\widetilde N$ and $\widetilde
N^\dagger$, we obtain the following four time-integrated decay rates
(in arbitrary units):
\beqa\label{decrat}
\Gamma(\widetilde L)&=&{\cal N}_s
|A_{\widetilde L}|^2\left[
  \frac{(1+|p/q|^2)(1+|\lambda_{\widetilde L}|^2)}{2(1-y^2)}
  +\frac{(1-|p/q|^2)(1-|\lambda_{\widetilde L}|^2)}{2(1+x^2)}\right.\no\\
&&+\left.\frac{y(1+|p/q|^2){\cal R}e\lambda_{\widetilde L}}{1-y^2}
  -\frac{x(1-|p/q|^2){\cal I}m\lambda_{\widetilde L}}{1+x^2}\right],\no\\
\Gamma(\widetilde L^\dagger)&=&{\cal N}_s|\bar A_{\widetilde L^\dagger}|^2
\left[\frac{(1+|q/p|^2)(1+|\lambda_{\widetilde L^\dagger}|^{-2})}{2(1-y^2)}
  +\frac{(1-|q/p|^2)(1-|\lambda_{\widetilde L^\dagger}|^{-2})}{2(1+x^2)}\right.\no\\
&&+\left.\frac{y(1+|q/p|^2){\cal R}e\frac{1}{\lambda_{\widetilde
L^\dagger}}}{1-y^2}
  -\frac{x(1-|q/p|^2){\cal I}m\frac{1}{\lambda_{\widetilde L^\dagger}}}{1+x^2}\right],\no\\
\Gamma(\overline L)&=&{\cal N}_f|A_{\overline L}|^2\left[
\frac{(1+|p/q|^2)(1+|\lambda_{\overline L}|^2)}{2(1-y^2)}
+\frac{(1-|p/q|^2)(1-|\lambda_{\overline L}|^2)}{2(1+x^2)}\right.\no\\
&&+\left.\frac{y(1+|p/q|^2){\cal R}e\lambda_{\overline L}}{1-y^2}
-\frac{x(1-|p/q|^2){\cal I}m\lambda_{\overline L}}{1+x^2}\right],\no\\
\Gamma(L)&=&{\cal N}_f|{\overline A}_L|^2\left[
  \frac{(1+|q/p|^2)(1+|\lambda_L|^{-2})}{2(1-y^2)}
  +\frac{(1-|q/p|^2)(1-|\lambda_L|^{-2})}{2(1+x^2)}\right.\no\\
&&+\left.\frac{y(1+|q/p|^2){\cal R}e\frac{1}{\lambda_L}}{1-y^2}
  -\frac{x(1-|q/p|^2){\cal I}m\frac{1}{\lambda_L}}{1+x^2}\right].
\eeqa
Using these four decay rates, we can obtain an exact expression for
$\varepsilon_\ell$ defined in eq. (\ref{defeel}).

%%%%%%%%%%%%%%%%%%
\section{The SSM+N}
\label{sec:ssm}
Since we are interested in the effects of the soft supersymmetry
breaking couplings, we work in a simplified single generation
model. The relevant superpotential terms are
\beq\label{suppot}
W=Y\epsilon_{\alpha\beta}L_{\alpha}N H_\beta+\frac12 MNN,
\eeq
where $L$ is the supermultiplet containing the left-handed lepton
doublet fields, $N$ is the superfield whose left-handed fermion is the
$SU(2)\times U(1)$-singlet ${\overline\nu}_L$, and $H$ is the Higgs
doublet (usually denoted by $H_2$). The relevant soft supersymmetry
breaking terms in the Lagrangian are the following:
\beq\label{lagssb}
{\cal L}_{\rm
  SSB}=-\left(m_2\lambda_2^a\lambda_2^a+A\epsilon_{\alpha\beta}\widetilde
  L_\alpha \widetilde NH_\beta+B\widetilde N\widetilde N+{\rm h.c.}\right).
\eeq
Here $\lambda_2^a$ ($a=1,2,3$) are the $SU(2)_L$ gauginos, $\widetilde
N,\widetilde L,H$ are scalar fields (and $N,L,h$ are their fermionic
superpartners). The $U(1)_Y$ gaugino, $\lambda_1$, would give effects
that are similar to those of $\lambda_2$ and can be included in a
straightforward way. 

The Lagrangian derived from eqs. (\ref{suppot}) and (\ref{lagssb}) has
two independent physical CP violating phases:
\beqa\label{cpvpha}
\phi_N&=&\arg(AMB^*Y^{*}),\no\\
\phi_W&=&\arg(m_2 MB^*).
\eeqa
These phases give the CP violation that is necessary to
dynamically generate a lepton asymmetry. If we set the lepton number
of $N$ and $\widetilde N$ to $-1$, so that $Y$ and $A$ are lepton
number conserving, the two couplings $M$ and $B$ violate lepton
number by two units. Thus processes that involve $Y$ or $A$, and
$M$ or $B$, would give the lepton number violation that is
necessary for leptogenesis.

There are several dimensionful parameters in (\ref{suppot}) and
(\ref{lagssb}). Of these $M$ is supersymmetry conserving and all other
are supersymmetry breaking. We assume the following hierarchies:
\beq\label{epssma}
\epss\equiv\frac{m_{\rm SUSY}}{M}\ll1,
\eeq
where $m_{\rm SUSY}$ is the supersymmetry breaking scale in the SSM+N
(we take $m_{\rm SUSY}\sim1$ TeV), and, unless otherwise stated,
\beq\label{asshie}
|m_2|\sim |A/Y|\sim |B/M| \sim m_{\rm SUSY}.
\eeq
We also assume that $|Y|\ll1$, as is required by the condition of
out-of-equilibrium decay [see eq. (\ref{uppyuk})].

We can evaluate the various parameters of eq. (\ref{decrat}) in terms
of the Lagrangian parameters of eqs. (\ref{suppot}) and
(\ref{lagssb}). The singlet sneutrino decay width is given, for
$|MY|\gg|A|$, by 
\beq
\Gamma=\frac{|MY^2|}{4\pi}.
\eeq
For the mixing parameters, we obtain
\beqa\label{xyqp}
x&=&\frac{2|B|}{|M|\Gamma}=\frac{8\pi|B|}{|MY|^2},\no\\
y&=&\left|\frac{A}{MY}\right|\cos\phi_N-\left|\frac{B}{M^2}\right|,\no\\
\left|\frac qp\right|&=&\left(1+\frac{2|AMY/(4\pi B)|\sin\phi_N}{1-|AMY/(4\pi
    B)|\sin\phi_N+\frac14|AMY/(4\pi B)|^2}\right)^{1/4}.
\eeqa

\begin{figure}[t]
\centerline{\includegraphics[width=0.7\textwidth]{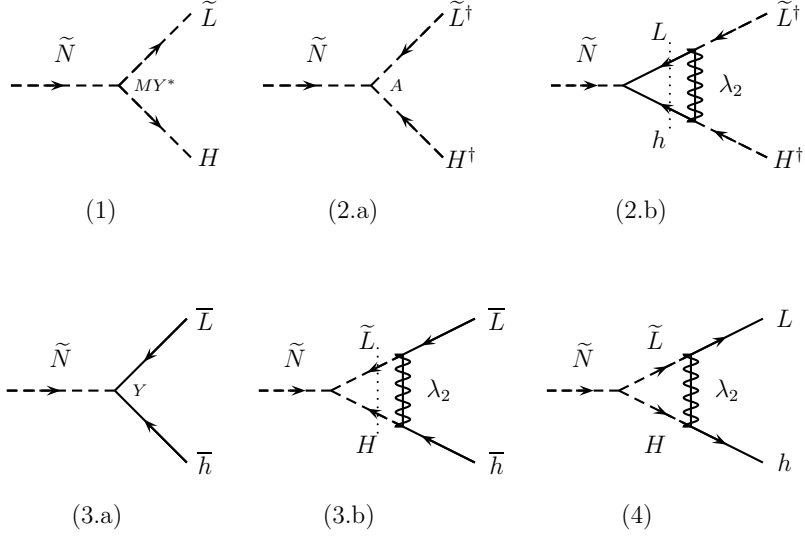}}
\caption{Two-body decay diagrams of a singlet sneutrino.}
\label{fg:twbo}
\end{figure}

As concerns the decay amplitudes, CPT guarantees the following relation:
\beq
|A_{\widetilde L}|^2+|A_{\widetilde L^\dagger}|^2+|A_{\overline
  L}|^2+|A_L|^2=
|{\overline A}_{\widetilde L}|^2+|{\overline A}_{\widetilde
  L^\dagger}|^2+|{\overline A}_{\overline L}|^2+|{\overline A}_{L}|^2.
\eeq
We consider only two body final states, since three body (or higher) 
states give only small corrections, as shown in Appendix
\ref{app:thr}. In Fig. \ref{fg:twbo} we show the relevant diagrams
(including the dominant one loop corrections) for the four two
body final states: (1) $\widetilde LH$, (2) $\widetilde L^\dagger
H^\dagger$, (3) $\overline L\overline h$, and (4) $Lh$. The amplitudes
are given by (we use an $=$ sign when the difference between the
absolute values of two CP conjugate amplitudes is negligible and an
$\approx$ sign when it is not)
\beqa
|A_{\widetilde L}|&=&|{\overline A}_{\widetilde
  L^\dagger}|=|MY|,\no\\
|A_{\overline L}|&\approx&|{\overline A}_{L}|\approx|MY|,\no\\
|{\overline A}_{\widetilde L}|&\approx&
|A_{\widetilde L^\dagger}|\approx|A|,\no\\
|{\overline A}_{\overline L}|&=&  
|A_L|=\frac{3\alpha_2}{4}|m_2Y|\sqrt{f_1^2+f_2^2}=
  {\cal O}\left(\alpha_2 |m_2 Y|\right),\no\\
|A_{\overline L}|&-&|{\overline A}_{L}|=-\frac{3\alpha_2}{2}
\left|\frac{m_2A}{M}\right|f_1\sin(\phi_W-\phi_N)
={\cal O}(\alpha_2 |m_2 A/M|),\no\\
|{\overline A}_{\widetilde L}|&-&
|A_{\widetilde L^\dagger}|=\frac{3\alpha_2}{2}
|m_2Y|f_1\sin(\phi_W-\phi_N)=
{\cal O}\left(\alpha_2 |m_2 Y|\right),
\eeqa
where $\alpha_2=g_2^2/(4\pi)$ is the weak coupling constant and where
we define
\beqa
f_1&=&\ln\left(\frac{M^2+m_2^2}{m_2^2}\right),\no\\
f_2&=&{\rm Li}_2\left(\frac{2}{1-\sqrt{1+4m_2^2/M^2}}\right)
+{\rm Li}_2\left(\frac{2}{1+\sqrt{1+4m_2^2/M^2}}\right).
\eeqa
These expressions assume, for simplicity, that $m_2>m_{\widetilde
  L},m_H$, thus neglecting corrections proportional to $m_{\widetilde
  L}$ and $m_H$. The function ${\rm
  Li}_2(z)\equiv\int_z^0\frac{\ln(1-t)dt}{t}$
is the dilogarithm function. For the relevant strong and weak phases,
we obtain 
\beqa
\phi_s&\equiv&\frac12\arg\left(\lambda_{\widetilde
    L}\lambda_{\widetilde L^\dagger}\right)=-\phi_N,\no\\
\phi_f&\equiv&\frac12\arg\left(\lambda_{\overline
    L}\lambda_L\right)=-\phi_W,\no\\ 
\sin\delta_s&\equiv&\sin\frac{\arg(\lambda_{\widetilde
  L}\lambda_{\widetilde L^\dagger}^{-1})}{2}=
\frac{3\alpha_2}{4}\left|\frac{m_2Y}{A}\right|f_1
={\cal O}(\alpha_2),\no\\
\sin\delta_f&\equiv&\sin\frac{\arg(\lambda_{\overline
  L}\lambda_L^{-1})}{2}=\frac{
f_1}{\sqrt{f_1^2+f_2^2}}
={\cal O}(1).
\eeqa

Note that there are several relations between decay amplitudes to
final scalars and to final fermions. These relations have to be taken
into account when evaluating the asymmetry. First, in the
supersymmetric limit we have $|A_{\widetilde
  L}|=|A_{\overline L}|$. Second, we have $|{\overline A}_{\widetilde L}
/A_{\widetilde L}|\sin\delta_s=|{\overline A}_{\overline L}/A_{\overline
  L}|\sin\delta_f$. Similar relations hold for the CP conjugate amplitudes.

%%%%%%%%%%%%%%%%%%
\section{The Leading Contributions to $\varepsilon_\ell$}
\label{sec:cpa}
Many terms that contribute to the lepton asymmetry are small and can
be neglected. The small parameters that play a role are the ratio
$\epss$, the weak coupling constant $\alpha_2$, and the Yukawa
coupling $Y$. The dependence on the Yukawa coupling enters either via
the combination $|A|/|MY|$ which, as can be seen from
eq. (\ref{asshie}), is taken to be of order $\epss$, or via the $x$
parameter evaluated in eq. (\ref{xyqp}). The $x$ parameter can be
small or large but, since $x\sim8\pi\epss/|Y|^2$, we take
$x\gg\epss$. (Some of the contributions that we consider are
significant only for $B\ll\epss M^2$ and, consequently,
$x\ll8\pi\epss/|Y|^2$. In these cases, however, $x\sim1$ is required,
so that $x\gg\epss$ is still valid.) We keep the $x$ dependence explicit.   

We identify several interesting contributions to
$\varepsilon_\ell$. We write down only the potentially leading
contributions and neglect terms that are suppressed by higher powers of  
$\epss$ and/or $\alpha_2$. We classify the contributions according to
the source of CP violation: 

(i) CP violation in mixing: Here, CP violation comes from
$|q/p|\neq1$ (as in R$e(\epsilon)$ in $K\to\pi\ell\nu$). We identify
two potentially significant contributions. The first is given by
\beq\label{epsmix}
\varepsilon_{1}^m=\frac{x^2}{4(1+x^2)}
\left(\left|\frac pq\right|^2 -\left|\frac qp\right|^2\right)\Delta_{sf}
={\cal O}\left(\frac{x\Delta_{sf}\epss}{1+x^2}\right).
\eeq
This is the contribution discussed in
\cite{Grossman:2003jv,D'Ambrosio:2003wy}. 
The size of this contribution depends crucially on
\beq\label{defdsf}
\Delta_{sf}\equiv \frac{{\cal N}_s(|A_{\widetilde L}|^2+|{\overline A}_{\widetilde
    L^\dagger}|^2)-{\cal N}_f( |A_{\overline L}|^2+|{\overline A}_L|^2)}
{{\cal N}_s(|A_{\widetilde L}|^2+|{\overline A}_{\widetilde
    L^\dagger}|^2)+{\cal N}_f(|A_{\overline L}|^2+|{\overline A}_L|^2)}.
\eeq
At zero temperature, $\Delta_{sf}={\cal O}(\epss^2)$, but for
temperature at the time of decay that is comparable to the singlet
sneutrino mass, $T_d\sim M$, we have $\Delta_{sf}\approx({\cal
  N}_s-{\cal N}_f)/({\cal N}_s+{\cal N}_f)={\cal O}(1)$.
The second contribution is given by (neglecting now corrections of
order $\Delta_{sf}$)
\beqa
\varepsilon_{2}^m&=&-\frac{x}{4(1+x^2)}\left(\left|\frac
    qp\right|-\left|\frac pq\right|\right)\left[\left(\left|\frac
    {\overline A_{\widetilde L}}{A_{\widetilde
        L}}\right|+\left|\frac{A_{\widetilde L^\dagger}}{{\overline
        A}_{\widetilde L^\dagger}}\right|\right)\sin\delta_s\cos\phi_s
-\left(\left|\frac
    {\overline A_{\overline L}}{A_{\overline
        L}}\right|+\left|\frac{A_L}{{\overline
        A}_L}\right|\right)\sin\delta_f\cos\phi_f\right]\no\\
&=&{\cal O}\left(\frac{\epss^2\alpha_2}{1+x^2}\right).
\eeqa

(ii) CP violation in interference of decays with and without mixing:
Here, CP violation comes from $\arg(\lambda_X\lambda_{\overline
  X})\neq0$ (as in $S_{\psi K_S}$ in $B\to J/\psi K$ and similar to
the mixing contribution to standard leptogenesis (see {\it e.g.}
\cite{Covi:1996fm}), though mixing in the latter case is between
different generations rather than between CP conjugate states). We
identify the following potentially significant contribution:
\beq\label{epsint}
\varepsilon^i=-\frac{y}{2}\left[\left(\left|\frac
    {\overline A_{\widetilde L}}{A_{\widetilde
        L}}\right|+\left|\frac{A_{\widetilde L^\dagger}}{{\overline
        A}_{\widetilde
        L^\dagger}}\right|\right)\sin\delta_s\sin\phi_s
-\left(\left|\frac
    {\overline A_{\overline L}}{A_{\overline
        L}}\right|+\left|\frac{A_L}{{\overline
        A}_L}\right|\right)\sin\delta_f\sin\phi_f\right]
={\cal O}\left(\epss^2\alpha_2\right).
\eeq

(iii) CP violation in decay: Here, CP violation comes from $
|A_X|\neq|{\overline A}_{\overline X}|$ (as in R$e(\epsilon^\prime)$
in $K\to\pi\pi$ and as in the vertex contribution to standard
leptogenesis). We identify the following potentially significant
contribution: 
\beq\label{epsdec}
\varepsilon^d=\frac{y}{2}\left(\left|\frac
    {\overline A_{\widetilde L}}{A_{\widetilde
        L}}\right|-\left|\frac{A_{\widetilde L^\dagger}}{{\overline
        A}_{\widetilde L^\dagger}}\right|\right)\cos\delta_s\cos\phi_s
={\cal O}\left(\epss^2\alpha_2\right).
\eeq

(iv) We also find a contribution that involves all three types of CP
violation and is not necessarily sub-dominant:
\beq
\varepsilon^{mdi}=-\frac{x}{4(1+x^2)}\left(\left|\frac
    qp\right|-\left|\frac pq\right|\right)\left(\left|\frac
    {\overline A_{\widetilde L}}{A_{\widetilde
        L}}\right|-\left|\frac{A_{\widetilde L^\dagger}}{{\overline
        A}_{\widetilde L^\dagger}}\right|\right)\cos\delta_s\sin\phi_s
={\cal O}\left(\frac{\epss^2\alpha_2}{1+x^2}\right).
\eeq

We note that, apart from $\varepsilon_1^m$, all the contributions
involve loop diagrams with gaugino exchange. The gaugino is playing a
double role here. First, its mass provides a new physical CP violating
phase. Second, the loop diagrams provide a strong phase. Consequently,
direct CP violation becomes a possible source of the lepton
asymmetry. Gaugino interactions do not violate lepton number, but they
allow the lepton number violating time evolution of the heavy
sneutrinos to contribute to $\varepsilon_\ell$ in new ways.
Without gaugino interactions, indirect CP violation is the
only significant source of soft leptogenesis
\cite{Grossman:2003jv,D'Ambrosio:2003wy}. Direct CP violation can
still be induced, but it involves higher powers of the Yukawa
couplings and is therefore negligibly small.

%%%%%%%%%%%%%%%%%
\section{The Size of $\varepsilon_\ell$}
\label{sec:siz}
In the previous section, we distinguished five potentially important
contributions to $\varepsilon_\ell$. These five contributions can be
separated into three different classes:
\beqa
\label{clasott}
\varepsilon_\ell&=&\varepsilon_1^m+(\varepsilon^i+\varepsilon^d)
+(\varepsilon_2^m+\varepsilon^{mdi}),\no\\
\varepsilon_1^m&=&{\cal
  O}\left(\frac{x\Delta_{sf}\epss}{1+x^2}\right),\no\\
\varepsilon^i,\varepsilon^d&=&{\cal
  O}\left(\epss^2\alpha_2\right),\no\\
\varepsilon_2^m,\varepsilon^{mdi}&=&{\cal O}\left(
  \frac{\epss^2\alpha_2}{1+x^2}\right).
\eeqa

The generated baryon to entropy ratio is given by
\beq
n_B/s\simeq-\kappa 10^{-3}\varepsilon_\ell,
\eeq
where $\kappa\lsim1$ is a dilution factor which takes into account the
possible inefficiency in the production of the heavy sneutrinos or
erasure of the generated asymmetry by lepton number violating
scattering processes. Since observations determine
$n_B/s\sim10^{-10}$, any of the contributions in (\ref{clasott}) would
be significant only if it yields 
$|\varepsilon_\ell|\gsim10^{-6}$. We now specify the conditions on the
parameters whereby each of the three classes of contributions can be
responsible for a successful leptogenesis. Since all the effects that
we consider are related to supersymmetry breaking and therefore
suppressed by powers of $\epss$, soft leptogenesis can give
significant effects only for $\epss\gsim10^{-6}$, that is,
\beq\label{uppermm}
M\lsim10^6\ m_{\rm SUSY}\sim10^9\ GeV.
\eeq

In order that the singlet neutrino and sneutrino decay out of
equilibrium, we should have a decay rate, $\Gamma=M|Y|^2/4\pi$, that
is not much faster than the expansion rate of the Universe,
$H=1.66g_*^{1/2}T^2/m_{\rm Pl}$ ($g_*$ counts the effective number of
spin degrees of freedom in thermal equilibrium; $g_*=228.75$ in the
SSM), at the time when the temperature is of order $M$: 
\beq\label{ouofeq}
M/|Y|^2\gsim3\times10^{16}\ GeV.
\eeq
On the other hand, the sneutrino decay should occur before the
electroweak phase transition, when sphalerons are still active,
$\Gamma>H(T\sim100\ GeV)$:
\beq\label{beewpt}
M|Y|^2\gsim 3\times10^{-13}\ GeV.
\eeq
Combining eqs. (\ref{uppermm}), (\ref{ouofeq}) and (\ref{beewpt}), we
learn that soft leptogenesis can give significant effects only for
\beq\label{uppyuk}
10^{-11}\ \left(\frac{10^9\ GeV}{M}\right)^{1/2}\lsim\
|Y|\lsim10^{-4}\ \left(\frac{M}{10^9\ GeV}\right)^{1/2}.
\eeq
With such a small Yukawa coupling, the decay width is rather narrow,
\beq
\Gamma\lsim1\ GeV\ \left(\frac{M}{10^9\ GeV}\right)^2.
\eeq

\begin{figure}[tb]
  \centering
  %\psfrag{LogY}{$\log(Y)$}
  %\psfrag{LogBonMm}{$\log(B/Mm_{SUSY})$}
  {\includegraphics[width=0.65\textwidth]{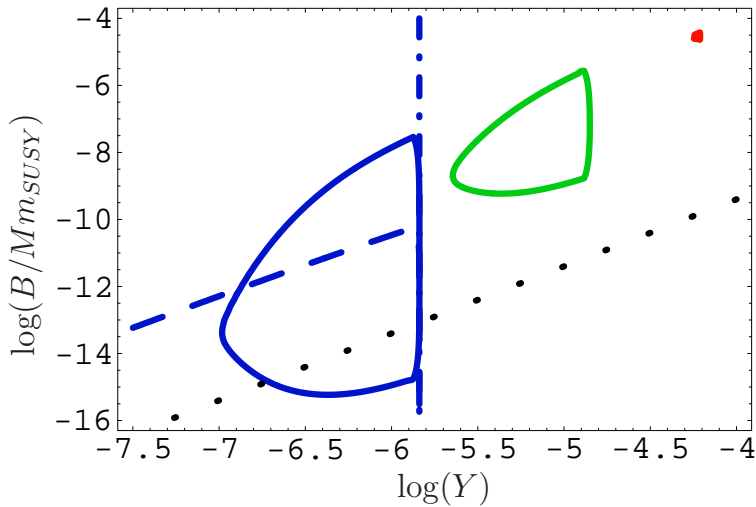}}
  \caption{Regions in the $Y-B$ plane where $\Gamma<H(T=M)$ and 
    $\varepsilon_\ell\gsim10^{-6}$. We take $m_{\rm SUSY}=10^3$
  GeV. The approximation made in our calculations ($x>\epss$) does not
  hold below the dotted line. (i) For
  $\varepsilon_\ell\sim\varepsilon_1^m$, the allowed 
  regions are within the solid curves, for $M= 2\times10^8$ (right),
  $10^7$ (middle) and $10^5$ (left) GeV. (ii) For
  $\varepsilon_\ell\sim\varepsilon^d+\varepsilon^i$, the allowed region
  is to the left of the dash-dotted line, with $M=10^5$ GeV. (iii) For
  $\varepsilon_\ell\sim\varepsilon^m_2+\varepsilon^{mdi}$, the allowed
  region is below and to the the left of the dashed curve,
  for $M=10^5$ GeV.}
  \label{fg:BY}
\end{figure}

(i) The contribution from $\varepsilon_1^m$ is of order
$(x/(1+x^2))\Delta_{sf}\epss$. For temperatures well below the mass
$M$, the finite temperature contribution to $\Delta_{sf}$ is given by
the following approximation ($n_{s,f}=(e^{M/(2T)}\mp1)^{-1}$):
\beq
\Delta_{sf}\simeq\frac{(1+n_s)^2-(1-n_f)^2}
{(1+n_s)^2+(1-n_f)^2}\approx2e^{-M/(2T_d)},
\eeq
where $T_d$ is the temperature at the time of decay.
To obtain $|\varepsilon_\ell|\gsim 10^{-6}$ we must have
\beq
\frac {T_d}{M}\gsim\frac{1}{2\ln(2\epss/10^{-6})}.
\eeq
By using $\Gamma=H(T_d)$, this can be translated into an upper bound
on $M/|Y|^2$: 
\beq\label{lardsf}
M/|Y|^2\lsim4\times10^{16}\ GeV[2\ln(2\epss/10^{-6})]^2.
\eeq
The lower (\ref{lardsf}) and upper (\ref{ouofeq}) bounds define, for
given $M$, a range for $|Y|$ and a range for $\Gamma$.
Finally, we must have
$x/(1+x^2)\gsim10^{-6}/(\Delta_{sf}\epss)$. Taking into
account that $x=2|B|/(M\Gamma)$, for a given value of $M$ we obtain an
allowed range for $B$. Since the naive estimate is $|B|\sim Mm_{\rm
  SUSY}$, it is useful to write the allowed range for $|B|$ in units of
$Mm_{\rm SUSY}$. We do so in
%Table \ref{tab:myg} and in
Fig. \ref{fg:BY}. We conclude that $\varepsilon_1^m$ can account for
the observed baryon asymmetry under the following conditions:
\begin{enumerate}
  \item The mass of the lightest sneutrino is light enough,
    $M\lsim10^9\ GeV$.
  \item The Yukawa couplings are small enough, $Y\lsim10^{-4}$. The
    lighter is $M$, the smaller the Yukawa coupling must be.
  \item The $B$ parameter is well below its naive value,
    $|B|/(Mm_{\rm SUSY})\lsim10^{-3}$. The lighter is $M$, the
    more suppressed the $B$ coupling must be.
  \end{enumerate}
We note that the inclusion of three body decays \cite{Allahverdi:2003tu}
does not change the basic picture and, in particular, does not modify
the estimate of $\Delta_{sf}$. We prove this statement in Appendix
\ref{app:thr}.

(ii) The contribution from $\varepsilon^d$ and $\varepsilon^i$ is of
order $\alpha_2\epss^2$. Since $\alpha_2\sim10^{-2}$, we must have
\beq\label{mtwocon}
M\lsim10^2 m_{\rm SUSY}.
\eeq
Eq. (\ref{uppyuk}) then requires
\beq\label{ytwocon}
Y\lsim10^{-6}.
\eeq
The region where this class of contributions can account for the
observed baryon asymmetry is to the left of the dash-dotted line in
Fig. \ref{fg:BY}. Note that $|B|$ is not constrained in this
scenario. In particular, it can take its naive value, $|B|\sim Mm_{\rm
  SUSY}$, in which case $x=2|B|/(M\Gamma)\gsim10^{11}$, so that the
sneutrino oscillation rate is much faster than its decay rate.

(iii) The contribution from $\varepsilon^m_2$ and $\varepsilon^{mdi}$ is of
order $\alpha_2\epss^2/(1+x^2)$. Consequently, the bound
(\ref{mtwocon}) on $M$ and the bound (\ref{ytwocon}) on $Y$ apply. In
addition, we must have $x\not\gg1$, which implies
\beq\label{bthrcon}
\frac{B}{Mm_{\rm SUSY}}\lsim\frac{M}{m_{\rm SUSY}}\ \frac{Y^2}{8\pi}\lsim10^{-11}.
\eeq
This third class of contributions is never much larger than the second
class. It may, however, be comparable if $B$ is small enough. The
region where this class of contributions is significant is to the left
and below the dashed line.

We note that, since our calculations are performed with the assumption
that $x\gg\epss$, they should not be trusted for $B/(Mm_{\rm
  SUSY})<Y^2/(8\pi)$, that is below the dotted line in
Fig. \ref{fg:BY}. 

%%%%%%%%%%%%%%%%
\section{Conclusions}
\label{sec:con}
Our main conclusions regarding the range of parameters where soft
leptogenesis may be successful are the following:
\begin{enumerate}
\item Soft leptogenesis can be neglected for $M\gg10^9$ GeV.
  \item Soft leptogenesis can work for $M\gg10^5$ GeV only if the
    Yukawa couplings have small values in a rather narrow range and
    if the $B$ parameter is very small compared to its naive scale
    ($Mm_{\rm SUSY}$). (We comment on the possibility of {\it
      naturally} achieving $B\ll Mm_{\rm SUSY}$ in Appendix
    \ref{app:smb}.) 
    \item For $M\lsim10^5$ GeV there are several
      contributions from soft leptogenesis that could account for the
      observed baryon asymmetry. All the supersymmetry soft breaking
      terms can assume their natural values.
      \end{enumerate}

The main novel point of this paper is the realization that soft
supersymmetry breaking terms give contributions to the lepton asymmetry
that are related to CP violation in decays [$\varepsilon^d$ of
eq. (\ref{epsdec})] and in the interference of decays with and
without mixing [$\varepsilon^i$ of eq. (\ref{epsint})]. In contrast to
CP violation in mixing [$\varepsilon^m$ of eq. (\ref{epsmix})], the
oscillation rate needs not be comparable to the decay rate in order to
have a significant effect. This is the reason that the $B$ term can
assume natural values. The new contributions to $\varepsilon_\ell$ are
second order in supersymmetry breaking terms and further suppressed by
a loop factor ($\sim\alpha_2 m_2A/(M^2Y)$) and are, therefore,
significant only if $M$ is not much higher than $10^2m_{\rm SUSY}$. 

The contribution to the lepton asymmetry related to CP violation in
mixing ($\varepsilon_1^m$ of eq. (\ref{epsmix})), which was originally
discussed in refs. \cite{Grossman:2003jv,D'Ambrosio:2003wy}, requires
thermal effects in order to be significant. In contrast, the new
contributions discussed here (such as $\varepsilon^i$ of
eq. (\ref{epsint}) and $\varepsilon^d$ of eq. (\ref{epsdec})) do not
require thermal effects and, consequently, allow a non-thermal
scenario of leptogenesis to work. Such a scenario would arise if, for
example, sneutrinos were produced by inflaton decays (or if the
sneutrino itself were the inflaton), and the temperature of the
thermal bath at the epoch of decay is well below $M$ (though above the
electroweak scale so that sphalerons are still active).

Soft leptogenesis opens up the interesting possibility that the scale
of the lightest singlet (s)neutrino mass ($M$) is not far above the
electroweak scale. In contrast, standard leptogenesis cannot yield, in
general, a
large enough asymmetry for low $M$. The difference between the two
scenarios lies in the different role of the Yukawa couplings. In both
standard and soft leptogenesis, the condition for out of equilibrium
decay associates a low scale $M$ with tiny Yukawa couplings $Y$. In
standard leptogenesis, the Yukawa couplings are the source of CP
violation; therefore, small $Y$ yield a small $\varepsilon_\ell$. In
soft leptogenesis, CP violation is induced by soft supersymmetry
breaking terms and is not suppressed by small $Y$.  
%%%%%%%%%%%
      
%%%%%%%%%%
\appendix
%%%%%%%%%%%%%%%%%%%

%%%%%%%%%%%%%%%%%%
\section{Three Body Decays}
\label{app:thr}
The $H_u$ field has Yukawa couplings to neutrinos and to up
quarks. The superpotential terms, $W=YNLH + Y_uQ\bar uH$, give a
quartic scalar interaction term in the Lagrangian,
\beq\label{lagfour}
{\cal L}_4=Y Y_u^{*}\widetilde N\widetilde L\widetilde Q^\dagger\widetilde{\bar
    u}^\dagger+{\rm h.c.} ,
\eeq
where $\widetilde Q$ is the scalar quark doublet and $\widetilde{\bar u}$
is the up-singlet.
This coupling allows the three body decay mode, $\widetilde N\to\widetilde
L^\dagger\widetilde{\bar u}\widetilde Q$. Since there is no similar quartic
coupling of $\widetilde N$ to two fermions and one scalar, one may think
that for the three body decays, the vanishing of $\Delta_{sf}^{(3)}$
(defined in (\ref{defdth})) in the supersymmetric limit is avoided, and
a sizeable lepton asymmetry is induced even at zero temperature. This
is, however, not the case, as we now explain.

\begin{figure}[t]
\centerline{\includegraphics[width=0.7\textwidth]{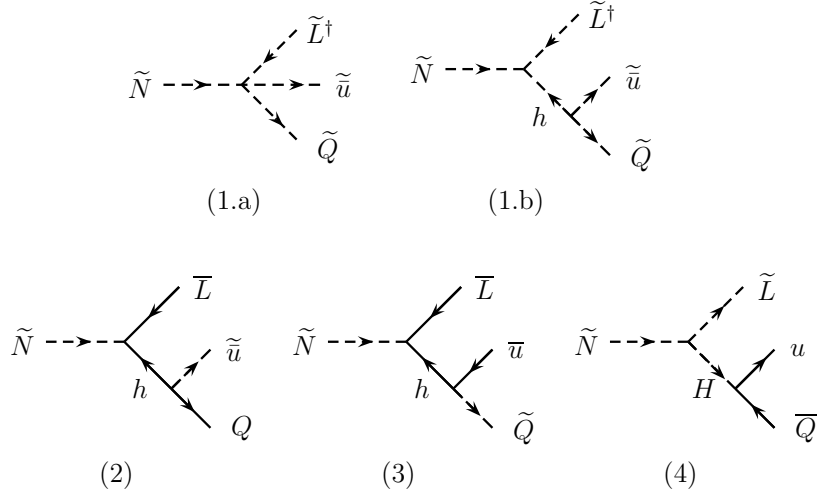}}
\caption{Three-body decay diagrams of a singlet sneutrino}
\label{fg:thbo}
\end{figure}

We are considering contributions to the CP asymmetry of the form
\beq\label{asythr}
\varepsilon^{(3)}=\frac{x^2}{4(1+x^2)}\left(\left|\frac qp\right|^2-
  \left|\frac pq\right|^2\right)\Delta_{sf}^{(3)},
\eeq
where
\beq\label{defdth}
\Delta_{sf}^{(3)}\equiv
\frac{\sum_{i=1}^4(-1)^{L_i+1}{\cal N}_i|A_i^{(3)}|^2}
{{\cal N}_s|{\overline A}_L|^2+{\cal N}_f|A_{\widetilde L}|^2}.
\eeq
Here $A_i^{(3)}$ is the amplitude of a relevant final three body state
with lepton number $L_i=\pm1$. At zero temperature, all the three body
phase space factors ${\cal N}_i$ are equal in the supersymmetric limit
and, consequently,
$\Delta_{sf}^{(3)}\propto\sum_{i=1}^4(-1)^{L_i+1}|A_i^{(3)}|^2$. The 
five tree level diagrams, leading to four different final states, are
shown in Fig. \ref{fg:thbo}. A straightforward calculation gives:
\beqa\label{fivethree}
\left|A_1^{(3)}\right|^2&=&\left|A_{1a}^{(3)}+A_{1b}^{(3)}\right|^2
=2|Y|^2|Y_u|^2\frac{m_{23}^4}{(m_{23}^2-\mu^2)^2},\no\\
\left|A_2^{(3)}\right|^2&=&2|Y|^2|Y_u|^2\frac{m_{12}^2 m_{23}^2}{(m_{23}^2-\mu^2)^2},\no\\
\left|A_3^{(3)}\right|^2&=&2|Y|^2|Y_u|^2\frac{m_{13}^2 m_{23}^2}{(m_{23}^2-\mu^2)^2},\no\\
\left|A_4^{(3)}\right|^2&=&2|Y|^2|Y_u|^2\frac{M^2 m_{23}^2}{(m_{23}^2-\mu^2)^2},
\eeqa
where $\mu$ is the supersymmetric mass of the $H_u$ supermultiplet,
and $m_{ij}^2=(k_i+k_j)^2$, with $k_1,k_2,k_3$ the momenta of,
respectively, the final (s)lepton, the singlet up (s)quark and
the doublet (s)quark. Then,
\beqa\label{asythra}
\sum_{i=1}^4(-1)^{L_i+1}|A_i^{(3)}|^2&=&|A_1^{(3)}|^2+|A_2^{(3)}|^2+|A_3^{(3)}|^2-|A_4^{(3)}|^2\no\\
&=&2|Y|^2|Y_u|^2\frac{m_{23}^2(M^2-m_{12}^2-m_{13}^2-m_{23}^2)}{(m_{23}^2-\mu^2)^2}=0.
\eeqa
The last equation, that is the vanishing of the
$\sum_{i=1}^4(-1)^{L_i+1}|A_i^{(3)}|^2$, holds in the supersymmetric limit,
when the three final particles are massless. The result is that, in
the supersymmetric limit, $\varepsilon_\ell^{(3)}=0$. The vanishing of
$\Delta_{sf}^{(3)}$ is lifted by finite temperature
effects, similarly to the case of $\Delta_{sf}$, but
then the contribution of the three body states is small compared to
the dominant two body ones.

If we assign lepton number $L=0$ to the $N$-supermultiplet, then 
the quantities $\Delta_{sf}$ defined in eq. (\ref{defdsf}) and
$\Delta_{sf}^{(3)}$ defined in eq. (\ref{defdth}) are the asymmetries
between $\Delta L=+1$ and $\Delta L=-1$ decay rates. Then, the
vanishing in the supersymmetric limit of $\Delta_{sf}$ and
$\Delta_{sf}^{(3)}$, demonstrated explicitly in our work, becomes
understandable on general grounds and generalizes to $n$-body
states for any $n$. In a single generation framework and in the
absence of supersymmetry breaking, singlet neutrino decay rates to
leptons and antileptons must be equal. Then, by supersymmetry, this
should hold also for singlet sneutrinos.

%%%%%%%%%%%%%
\section{On the naturalness of $B=0$}
\label{app:smb}
We consider the following superpotential terms:
\beq
W=MNN+YNLH,
\eeq
and SUSY breaking terms,
\beq
{\cal L}=B\widetilde N\widetilde N+A\widetilde N\widetilde LH.
\eeq
In the absence of these terms, there are four additional flavor
conserving global $U(1)$ symmetries: $U(1)_N\times U(1)_L\times
U(1)_H\times U(1)_R$, with the following charge assignments:
\beqa
N(1,0,0,0),&\ \ \ \  &\widetilde N(1,0,0,1),\no\\
L(0,1,0,0),&\ \ \ \  &\widetilde L(0,1,0,1),\no\\
h(0,0,1,0),&\ \ \ \  &H(0,0,1,1).
\eeqa
Selection rules for the symmetries may be used if $M,Y,A$ and $B$ are
treated as spurions with charges assigned to compensate those of the
fields:
\beq
M(-2,0,0,0),\ \
Y(-1,-1,-1,-1),\ \
A(-1,-1,-1,-3),\ \
B(-2,0,0,-2).
\eeq
To understand the consequences, it is simpler to examine the
charges of the spurions under $U(1)_{N-L}\times U(1)_{2R-3(L+H)}\times
U(1)_{2R-(L+H)}\times U(1)_{L+H}$:
\beq
M(-2,0,0,0),\ \
Y(0,+4,0,0),\ \
A(0,0,-4,0),\ \
B(-2,-4,-4,0).
\eeq
We learn that setting $B=0$ does not add a symmetry to the
Lagrangian. Consequently, $B$ is {\it additively} renormalized.
However, setting $B$ and any other of the three couplings,
$M,Y$ or $A$, to zero is natural.

We can therefore think of a three generation framework where, for
example, $Y=0$ because of a supersymmetric Froggatt-Nielsen symmetry
\cite{Froggatt:1978nt,Leurer:1993gy}. Then $B=0$ is
natural. When the FN symmetry is spontaneously broken, $B$ can be
naturally suppressed:
\beq\label{supb}
B\propto AMY^\dagger\ll AM.
\eeq
Of course, a Froggatt-Nielsen symmetry can also induce $A\sim
m_{\rm SUSY}Y\ll m_{\rm SUSY}$, leading to further suppression of $B$ compared to
$Mm_{\rm SUSY}$. Both the additive renormalization, and the suppression
factors in (\ref{supb}) are manifest in the RGE \cite{Martin:1993zk}:
\begin{equation}\label{brun}
16\pi^2\frac{d}{dt}B=MY^*A,
\end{equation}
If, however, $B$ is radiatively generated, as in (\ref{brun}), the 
phase $\phi_N$ vanishes at this order. At two loops, there will be a
contribution to $B$ that depends on $m_2$, but then
$\phi_N\sim\alpha_2$. 

%%%%%%%%%%%%
\acknowledgments
We thank Lance Dixon, Gian Giudice, Riccardo Rattazzi and Yael Shadmi
for useful discussions. This work was supported in part by a grant
from Fundaci\'on Antorchas/Weizmann.  The research of Y.G. and Y.N. is
supported by a Grant from the G.I.F., the German--Israeli Foundation
for Scientific Research and Development, and by the Einstein Minerva
Center for Theoretical Physics.  
The work of Y.G. is supported  by
the United States--Israel Binational Science Foundation through grant
No.~2000133, by the Israel Science Foundation under grant No.~237/01,
by the Department of Energy, contract DE-AC03-76SF00515 and by the
Department of Energy under grant No.~DE-FG03-92ER40689.
Y.N.~is supported by the Israel Science Foundation
founded by the Israel Academy of Sciences and Humanities, by EEC RTN
contract HPRN-CT-00292-2002, by the Minerva Foundation (M\"unchen),
and by a grant from the United States-Israel Binational Science
Foundation (BSF), Jerusalem, Israel. ER is partially supported by a
John Simon Guggenheim Foundation fellowship. 

%%%%%%%%%%%%%%%%%%%%%

\end{document}